# Alternative Effort-optimal Model-based Strategy for State Machine Testing of IoT Systems

Vaclav Rechtberger
Dept. of Computer Science, FEE
Czech Technical University in Prague
Karlovo nam. 13, 121 35 Praha 2
Czech Republic
rechtva1@fel.cvut.cz

Miroslav Bures
Dept. of Computer Science, FEE
Czech Technical University in Prague
Karlovo nam. 13, 121 35 Praha 2
Czech Republic
miroslav.bures@fel.cvut.cz

Bestoun S. Ahmed
Karlstad University
and FEE, CTU in Prague
Universitetsgatan 2, 65188, Karlstad
Sweden
bestoun@kau.se

## ABSTRACT
To effectively test parts of the Internet of Things (IoT) systems with a state machine character, Model-based Testing (MBT) approach can be taken. In MBT, a system model is created, and test cases are generated automatically from the model, and a number of current strategies exist. In this paper, we propose a novel alternative strategy that concurrently allows us to flexibly adjust the preferred length of the generated test cases, as well as to mark the states, in which the test case can start and end. Compared with an intuitive N-switch coverage-based strategy that aims at the same goals, our proposal generates a lower number of shorter test cases with fewer test step duplications.

## CCS Concepts
• **Software and its engineering** → **Software creation and management** → **Software verification and validation**

## Keywords
Regression Testing; Model-based Testing; State Machine; Test Case Generation; Test Automation.

## 1. INTRODUCTION
Model-based Testing (MBT) is commonly employed approach to testing of various software and electronic systems, or their hybrids, which are becoming more widespread and popular in the recent decade with the extensive development of the Internet of Things (IoT) systems. Especially in the IoT domain, the MBT approaches can answer to current reliability and security challenges reported in this field [1,2,3].

In MBT, we model the System Under Test (SUT) or its part (or selected aspect) by a suitable model, capturing relevant information about the SUT, which is needed for its testing. The model then typically serves as a basis for automated generation of the test scenarios covering the selected aspects of the SUT (although it is not the only possible form of MBT; more variants exist [4,5]).

For the parts of the SUT having a character of the state machine, various notations of state machines are used to model it for MBT purposes, spanning from a formal grammar-based definition of the automaton to models based on directed graphs (e.g., UML statechart diagram [6]). In this study, we focus on the directed graph-based model.

In a generation of the test cases automatically from such a model, the test coverage criteria concept is used. Generally speaking, the test coverage determines the extent to which the test cases exercise the model's elements. In the state machine testing, *the N-switch* coverage concept is used. To satisfy the *N-switch* coverage, the test cases must contain all possible sequences of $N+1$ consecutive transitions in the state machine [4,7]. As specific cases of *N-switch* coverage, state coverage (sometimes referred to as all-state coverage) or transition coverage are also employed [6,7,8], meaning that each state or transition in the SUT model has to be exercised at least once by the test cases.

For several types of test assignments, where the lower intensity of tests is required as end-user testing or regression testing, the test cases strictly based on *the N-switch* coverage criterion. This might be sub-optimal and a more dynamic test case generation strategy might yield better results from an overall testing economics viewpoint.

These potentially better results can be achieved by the combination of the following two elements:

1. We allow to limit the length of generated test cases due to test analyst's preferences, and,
2. we define possible starts and possible ends of the test cases in the model to allow the test analyst to simulate conditions possible, limiting the execution of the tests in a real project.

Considering the advantages of employing these both elements concurrently, we created a novel test case generation strategy, named Flexible State Machine Test (FSMT). This strategy is presented further in this paper.

The paper is organized as follows: Section 2 discusses the related work in the fields comparable to our proposal. Section 3 introduces the underlying SUT model and explains the proposed test case generation strategy. Section 4 presents and discusses the experimental results. The last section concludes the paper.

## 2. RELATED WORK
Several strategies have been explored to generate test cases for SUT parts having a character of a state machine, differing by particular test coverage criteria [8] that has to be satisfied in the generated test cases [5,6,7].

In this section, we select and discuss several proposals that are mostly close to our intention and approach. However, all state-of-the-art studies are different from our proposal in essential details, which we discuss in this section.

In optimization of test sets exercising the parts of a SUT having a character of a state machine, comparable examples include optimization of regression test sets based on the state machine-based SUT model [9], optimization of test cases using an approach based on basis path minimization [10], or generation of the state-machine based test cases using various modifications of genetic algorithms [11].

However, none of these proposals aims directly at the goal which we address in this study. For instance, a proposal by Tahat et al. focuses on the regression test set optimization, using a more significant reduction in the test set that our intention is in the proposed strategy [9]. A proposal by Anbunathan et al. follows a line of thought initially similar to our proposal. However, principal differences occur in the proposed strategy [10], as we allow to limit the length of the test cases and to explicitly select inner states of the SUT model as possible start and end of the test cases (see Section 3). The same difference is also relevant to the proposal by Turlea et al. [11].

From the recent works, the *N-switch* coverage concept is directly supported in the proposal by Pradhan et al. The concept is denoted as All Transition Pair (ATP), however, as the name might suggest 1-switch coverage, it is equivalent to general *N-switch* coverage [12]. However, the proposal does not explicitly allow the selection of inner states of the SUT model as possible start and end of the test cases. Also, adjustment of test case length is limited compared to our proposal that follows in this paper.

ATP criterion is also discussed in the study by Devroey et al., however, the algorithm provided in the paper focuses subsequently only on the all-state coverage; moreover, the differences as discussed above are also relevant for this study [13].

To our present knowledge, we have not found a state-of-the-art study that follows the same test case generation strategy, which we present in this paper.

## 3. PROPOSED STRATEGY
We explain the SUT model we are using for the FSMT strategy and its principle in the following subsections.

### 3.1 SUT Model
In the FSMT strategy, we employ the SUT model defined as a directed graph $G$ ($V$, $E$, $v_s$, $V_e$, $V_{ts}$, $V_{te}$), where vertices $V$ model states of the system and edges $E$ model transitions between the states. In the model, $v_s \in V$ is a start vertex of the state machine, $V_e \subseteq V$ is a set of end vertices of the state machine, $V_{ts} \subseteq V$ is a set of the possible start of test paths, and $V_{te} \subseteq V$ is a set of the possible end of test paths.

Further, $v_s \in V_{ts}$ and $V_e \subseteq V_{te}$ and $V_{ts}$ and $V_{te}$ can have nonempty intersect.

Test case $p$ is a path in $G$ that starts in a $v_{ts} \in V_{ts}$ and ends in a $v_{te} \in V_{te}$. We denote a test path as a sequence of edges. The output of the algorithm is a set of test paths $P$. Test path and test case have the synonymous meaning in this text.

As introduced above, in the test case generation strategy, we combine the following two elements.

(1) The N-switch coverage does not strictly drive the length of the test cases; instead of, it is dynamic and uses the minimal and maximal length bounds given by the test analyst (denoted as *minLength* and *maxLength*).

(2) The test cases might start and end only in defined states of the SUT state machine model $V_{ts}$ and $V_{te}$, which are truly relevant in the testing process: in some SUT states, starting and ending the real tests might be impossible due to the technical nature of the SUT.

Using the **G**, the test case generation strategy is formulated.

### 3.2 Test Case Generation Strategy
The FSMT strategy accepts SUT model **G**, minimal and maximal length of test paths (*minLength* and *maxLength*) and produces a set of test cases *P*.

FSMT generates the *P* using the following principal algorithm:

```
E_unc is a set of SUT model edges uncovered by
generated test cases, initially E_unc = E.

For each v ∈ V_ts do:
    Find the shortest path from v to a possible
    end state from V_te, length of which is in the
    interval from minLength to maxLength. If
    found, denote this path as p, and:
        Find out, which edges from E_unc have been
        covered by p and remove these edges from
        E_unc.
        Add p to P.
Until E_unc is not empty, do:
    Get a random edge e ∈ E_unc and remove e from
    E_unc.
    Find the shortest path that is going from a
    state from V_ts to e and then from e to a state
    of from V_te. Length of such a path must be in
    interval from minLength to maxLength. If
    found, denote this path as p, and:
        Find out, which edges from E_unc have been
        covered by p and remove these edges from
        E_unc.
        Add p to P.
```

At the end of the run of the algorithm, *P* contains created test cases, and $E_{unc}$ is empty. The nature of the problem guarantees that the algorithm stops deterministically.

### 3.3 Strategy Implementation
We implemented the FSMT strategy in the Oxygen MBT platform issued by our research group. The beta version of the platform with the implemented strategy is available for free public use. Oxygen is implemented in Java and runnable as an executable JAR file, requiring Java 1.8. The application can be download at http://still.felk.cvut.cz/download/oxygen3.zip.

In the Oxygen, the test analyst creates the SUT model **G** using a visual editor (a sample is provided in Figure 1).

Used visual schema is based on simplified UML notation; the only exception is a notation of the states belonging to $V_{ts}$ and $V_{te}$. If a state belongs to $V_{ts}$, the green background is used.

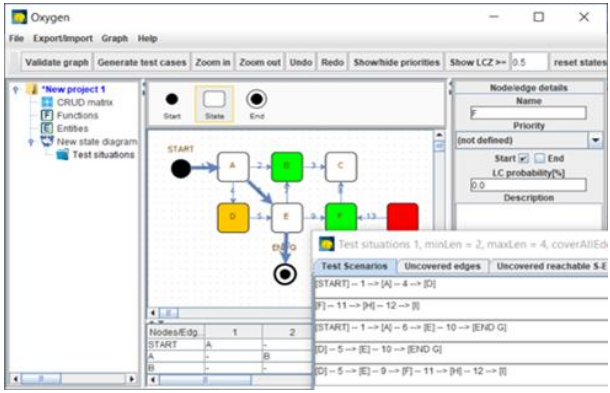

**Figure 1. SUT model creation in Oxygen platform.**

If a state belongs to $V_{te}$, the red background is used, and if a state belongs to both $V_{ts}$ and $V_{te}$, such a state is distinguished by a yellow background.

When the SUT model is created, test cases can be generated automatically using the strategy described in Section 3.2. Generated test cases can be visualized in the SUT model (an example is given in Figure 1, test case selected in the pop-up window is visualized in the SUT model by bold arrows).

Mainly, the created test cases can be exported in open formats based on XML, CSV, and JSON. These formats can be easily used to load the test case to a test management tool or test automation framework.

## 4. EXPERIMENTS

The proposed strategy is currently employed in several verification projects, including proof of concept in Skoda Auto automotive producer. In this project, the strategy is used to design test scenarios for end-used testing of produced automobiles.

In this section, we compare the proposed strategy with its alternative, based on *the N-switch* coverage concept, which is, due to our best knowledge, the most comparable candidate to evaluate the strategy. Considering the other state-of-the-art-strategies, we concluded, that comparability of the different algorithms published in the literature is limited, due to the nature of our strategy, enforcing defined length range of the test cases and their starts and ends in defined states of the SUT model.

### 4.1 Compared strategy

To concurrently employ both elements characteristic for the proposed FSMT strategy, which are (1) limitation of the length of generated test cases due to test analyst's preferences, and, (2) test cases starting and finishing only in defined states, the comparable strategy based on *N-switch* coverage concept needs to be as defined in the following principal algorithm:

```
Create all possible N-switch sequences for G where
N is in the interval from minLength-1 to maxLength-
1. Put these sequences to a set P.

Remove all sequences that are not starting in any
state from Vts from the P.

Remove all sequences that are not ending in any
state from Vte from the P.
```

Due to the nature of the compared strategy, we denote this *N-switch* based alternative as Brute Force Approach (BFA) for further reference.

### 4.2 Experiment method and set up

We implemented BFA in the benchmarking module of the Oxygen platform, which allows a comparison of the algorithms computing the test cases using one SUT model (we have applied this system successfully several times during the development of MBT algorithms instance [14]).

The benchmarking module allows multiple runs of the algorithm with various SUT models prepared in the Oxygen platform. The module records selected properties of SUT models as well as properties of the test sets created by individual compared algorithms (both mentioned properties are introduced further in this section). The results are then exported in CSV format for further processing and analysis.

In the experiment, we created 40 problem instances (different SUT models) that varied by $|V|$, $|E|$, number of cycles in $G$, the average length of these cycles, number of parallel edges in $G$, $|V_{ts}|$ and $|V_{te}|$.

In the SUT models, average $|V|$ was 17.3., ranging from 5 to 40. Average $|E|$ was 35.4, ranging from 10 to 80. The average number of cycles in the SUT models 6.5, three models having no cycles and 12 of models having more than ten cycles. The average length of cycles present in the used SUT models was 5.

The average number of parallel edges in the models was 6,8. Out of all models, 18 did not have any parallel edges, and six models had more than 20 parallel edges. The average node degree of all models was 4.4. Finally, average $|V_{ts}|$ in all models was 4.1, and average $|V_{te}|$ was 4.7. The average number of states, in which a test case can start as well as in which can end was 2.6.

In the experiment, we employed FSMT and BFA strategies to create *P* for the SUT models introduced above. We analyzed the properties of the generated test cases, including the number of test cases in *P*, the total number of test case steps, the average length of the test cases, and the ratio of unique edges in the test cases.

We performed these measurements for three sets of length range (*minLength* to *maxLength* interval), particularly 1 to 2, 3 to 4, and 1 to 4. The results are presented in the following subsection.

### 4.3 Results

Table 1 presents the measured properties of the test cases generated by FSMT and BFA, averaged for all SUT models used in the experiments. For individual *minLength* to *maxLength* ranges, the difference of values is also presented.

In Table 1, *length* denotes the total number of test case steps measured in a number of edges in a test case, $|P|$ denotes the number of test cases in a test set *P*, *avg length* denotes the average length of test cases in a test set *P* and *unique* denotes a ratio of the number of unique edges in *P* to all number of edges in *P*. The *lower* the unique value is, the more edge duplications the *P* contains.

For the set of SUT models used in this experiment, FSMT outperformed BFA practically in all measured parameters. For compared BFA, the total length of the test cases (*length*), which is the major indicator, determining the effort needed to execute the tests, was approximately two times higher for *minLength-maxLength* range 1 to 2 and approximately three times higher for ranges 3 to 4 and 1 to 4.

**Table 1. Properties of the generated tests cases**

| Strategy | length | |P| | avg length | unique |
|---|---|---|---|---|
| *minLength* = 1, *maxLength* = 2 | | | | |
| FSMT | 7.8 | 5.8 | 1.3 | 95.7% |
| BFA | 18.2 | 10.6 | 1.7 | 41.0% |
| *difference* | 2.3 | 1.8 | 1.3 | 0.4 |
| *minLength* = 3, *maxLength* = 4 | | | | |
| FSMT | 51.8 | 15.6 | 3.3 | 56.6% |
| BFA | 151.9 | 40.7 | 3.7 | 19.3% |
| *difference* | 2.9 | 2.6 | 1.1 | 0.3 |
| *minLength* = 1, *maxLength* = 4 | | | | |
| FSMT | 53.3 | 21.6 | 2.4 | 57.0% |
| BFA | 170.1 | 51.3 | 3.1 | 17.9% |
| *difference* | 3.2 | 2.4 | 1.3 | 0.3 |

The similar, but less significant trend could be observed for the total number of test cases (|P|), where for BFA, this number was approximately two times higher for *minLength-maxLength* range 1 to 2 and approximately two and half times higher for ranges 3 to 4 and 1 to 4.

A slight, but no significant difference was found for an average length of the test cases (*avg length*), in favor of the FSTM strategy.

Finally, test cases generated by the BFA strategy had a significantly higher *unique* ratio, indicating that duplication of the edges in these test cases is much higher than in the test cases generated by the FSMT strategy. This ratio is more than twofold higher for *minLength-maxLength* range 1 to 2 and approximately threefold higher for ranges 3 to 4 and 1 to 4.

Interpretation and further analysis of the results follow in the next subsection.

### 4.4 Discussion

The preliminary results show good results of the proposed FSMT strategy compared to comparable BFA strategy, based on the *N-switch* coverage concept and certain "brute force" approach. Such a strategy can be expected to be taken by a test analyst, without FSMT strategy support.

From the obtained results, it is obvious that strategy such as BFA is not optimal in achieving the goals of the FSMT, and a special strategy, as formulated in this paper, is needed to acquire a close-to-optimum set of test cases.

Unfortunately, no state-of-the-art strategy or algorithm better than BFA is directly comparable with the proposed FSMT due to its specific characteristics, flexible length of test cases limited to a given interval and definition of possible starts and ends of the test cases.

It is important to note that the comparison of the test sets in the presented experiment is based solely on the properties of the test cases, which hat to satisfy the mentioned goals of the FSMT. The situation has to be analyzed from the test's economic viewpoint and from the probability of the test cases to find a defect.

From the testing process economics viewpoint, FSMT generates a lower number of shorter test cases, which favors FSMT against BFA. Especially for manual testing, it can be expected that the total effort needed for tests correlates with the number of test steps. Hence, the test cases produced by the FSMT are nearer to an optimum from the total effort viewpoint.

However, the FSMT produces the tests with less test step duplications than BFA. Regarding the probability of the test cases to find a defect in this point, the duplications might, on the other hand, raise the likelihood of detecting some more defects than can be done by test cases produced by FSMT. It can be expected that such a gain in probability to detect more defects by test case duplications would not overweight gain in shorter test cases obtained by FSMT. More extensive experiments have to be done at this point, typically using a mutation testing or defect injection technique [15,16]. Getting more insight into this matter is part of our planned future work.

To conclude, it is essential to note that the described effect is entirely natural for the general test design process. More steps of the test cases (even duplicate ones) usually lead to a higher probability of detecting some defects. The goal is to find a minimal test set, which is still having the highest relative chance to detect the defects. From this viewpoint, FSMT aims at such optimization.

### 5. CONCLUSION

In this study, we proposed FSTM, a novel test case-generation strategy applicable to the parts of the SUTs with a state machine's character. In contrast to previous strategies, employing various test coverage criteria is usually derived from the *N-switch* concept. The FSMT applies two concurrent elements: possibility to determine a length range for the generated test cases flexibly and the possibility to select the states, in which a test case can start and end. These features give the FSMT potential to work more effectively in the practical testing process, reflecting the needs of test analysts for the expected length of test cases and reflecting the real situations, in which a test case can be started or ended in certain states of the system.

The FSMT is primarily aimed at testing various IoT systems, for which a character of a state machine is typical; however, the strategy is generally applicable to any software or electronic system.

As the experimental result show, compared to an intuitive strategy based on *N-switch* coverage concept (BFA), that is trying to achieve the same goal as FSTM, our proposal generates a lower number of shorter test cases with less test step duplications. Overall, the difference in the total length of test cases was two-fold up to three-fold, based on the expected length of test cases and was two-fold up to two-and-half-fold regarding the number of test cases. However, more experiments will be conducted to get further insight into the relative probability of the test cases produced by compared strategies to detect defects in a SUT, as more duplication of the steps in the test cases generated by BFA might increase its probability. This effect could potentially lower the outperformance of the proposed FSMT against the BFA.

### 6. ACKNOWLEDGMENTS

This study is conducted as a part of the project TACR TH02010296 „Quality Assurance for Internet of Things Technology". This work has been supported by the OP VVV funded project CZ.02.1.01/0.0./0.0./16_019/0000765 „Research Center for Informatics". We would like to also thank Vladimir Skopek and his colleagues from the Skoda Auto car testing team for their valuable feedback to the proposed strategy.